\documentclass{blois}

\def\be{\begin{equation}}
\def\ee{\end{equation}}
\def\bea{\begin{eqnarray}}
\def\eea{\end{eqnarray}}

\newcommand{\Photo}{\includegraphics[height=35mm]{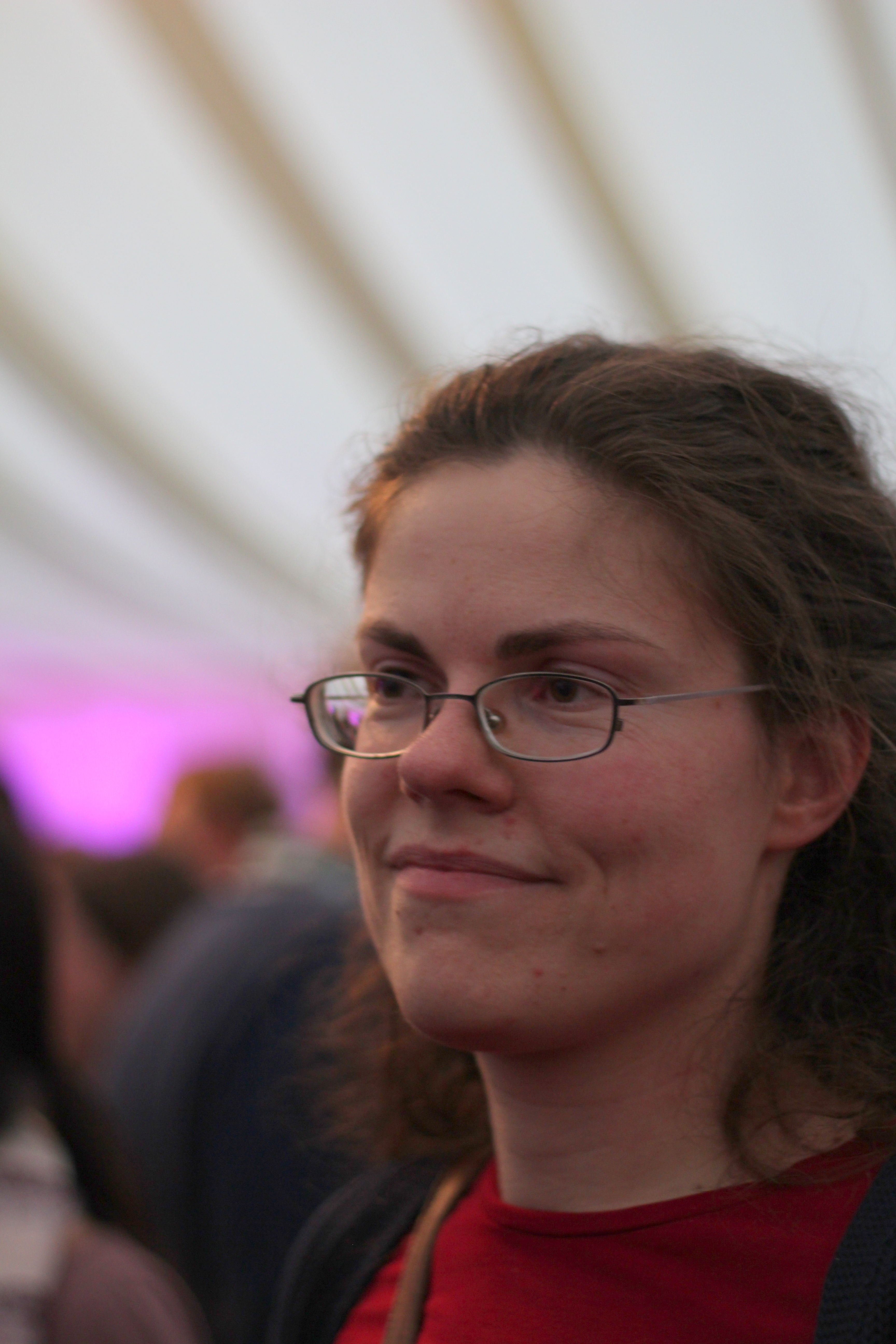}}

\begin{document}
\vspace*{4cm}
\title{Recent Results of Top Quark Physics from the Tevatron}

\author{Reinhild Yvonne Peters}

\address{University of Manchester, School of Physics and Astronomy,
  Oxford Road, Manchester M13 9PL, England; also at DESY, Hamburg, Germany }

\maketitle\abstracts{
Twenty years after its discovery  in 1995  by the
CDF and D0 collaborations at the Tevatron proton-antiproton collider
at Fermilab, the top
quark still undergoes intensive studies at the Tevatron
and the LHC at CERN. In this article, recent top quark physics results
from CDF and D0 are reported. In particular,
measurements 
of single top quark and double top quark production, the $t\bar{t}$
forward-backward asymmetry and the top quark mass are discussed.}

\section{Introduction}
The heaviest known elementary particle, the top quark, has been
discovered in 1995 by the CDF and D0 Collaborations in Run~I of the
Fermilab Tevatron proton-antiproton
collider~\cite{cdftopdiscovery}~\cite{d0topdiscovery}. Due to its high
mass, the top quark is believed to play a special role in electroweak
symmetry breaking, and is also considered a window to new physics in many models
beyond the standard model (SM). Furthermore, the top quark is the only quark that  allows to study a bare
quark. 

The results shown in this article are based on the full Tevatron
Run~II data sample,
collected at a collision energy of 1.96~TeV. Run~II started in 2001,
and ended on September 30th 2011, providing $\approx 10.5$~fb$^{-1}$ of integrated luminosity for
each of 
the D0 and CDF experiments. 

In this article, latest measurements in the single top sector, results
of top antitop quark  ($t\bar{t}$) production, the $t\bar{t}$
forward-backward asymmetry, and the top quark mass, are discussed.

\section{Single Top}
Top quark production dominantly occurs in pairs via the strong
interaction, or singly via the electroweak interaction. Single top
quark production happens via  $s$-channel, $t$-channel
and $Wt$-channel processes.  The latter has a negligible cross section at the
Tevatron.
 
The measurement of the single top quark production cross section is
quite challenging, since the production cross section for the main
background from $W$+jets processes is orders of magnitude larger and
has a very similar final state to single
top. Various multivariate analysis techniques are used, combining
multiple variables into a discriminant. The first observation of single top
quark production was achieved in 2009 by CDF and
D0~\cite{cdfsingletop}~\cite{d0singletop}, with $s$- and $t$-channels
combined. The observation was based on $3.2$~fb$^{-1}$ and
$2.3$~fb$^{-1}$ of data by CDF and D0, respectively. In 2011, single
top $t$-channel production has been  first observed by D0~\cite{d0tchannel} using 5.4~fb$^{-1}$ of data. 
Finally, in 2014, also the $s$-channel production was observed by
combining the CDF and D0 measurements~\cite{tevschannel}, using
analyses which are based on up to the full Run~II data sample. Semileptonic
events are considered in the analyses by both collaborations, with the
addition of an analysis by CDF, where 
events with a missing transverse energy plus jet signature are used,
adding events to the sample in which the lepton is not directly reconstructed. The combined analysis results in a cross section of
$\sigma_{s}=1.29^{+0.26}_{-0.24}$~pb, which deviates with more than  6.3 standard
deviations (SD) from zero. 
Recently, a final Tevatron combination of the single top quark cross
sections has been performed. This comprises a combined measurement of the
$t$-channel cross section, $\sigma_{t}= 2.25^{+0.29}_{-0.31}$~pb, a
two-dimensional measurement of the $s$- versus $t$-channel cross
sections, and a measurement of  the $s+t$-channel cross section,
$\sigma_{s+t}=3.30^{+0.52}_{-0.40}$~pb~\cite{tevcombinew}. Using the
single top cross section, the CKM matrix element $|V_{tb}|$ can be
extracted. The measured value of  $|V_{tb}| = 1.02^{+0.06}_{-0.05}$
corresponds to $|V_{tb}|>0.92$ at the 95\% C.L.

\section{Double Top}
Top quark pair production occurs via $q\bar{q}$ annihilation  or
gluon-gluon fusion. At the Tevatron, the former comprises
approximately 85\% of $t\bar{t}$ production, and the latter about
15\%, which is roughly the other way round at the LHC. Besides the
different collision energies, these different fractions are one of the
main reasons why many measurements at the Tevatron are complementary to
similar studies at the LHC. 

\begin{figure}[h]                                                                                                                                                                                                                              
\begin{minipage}{18pc}                                                                                                                                                                                                                         
\includegraphics[width=20pc]{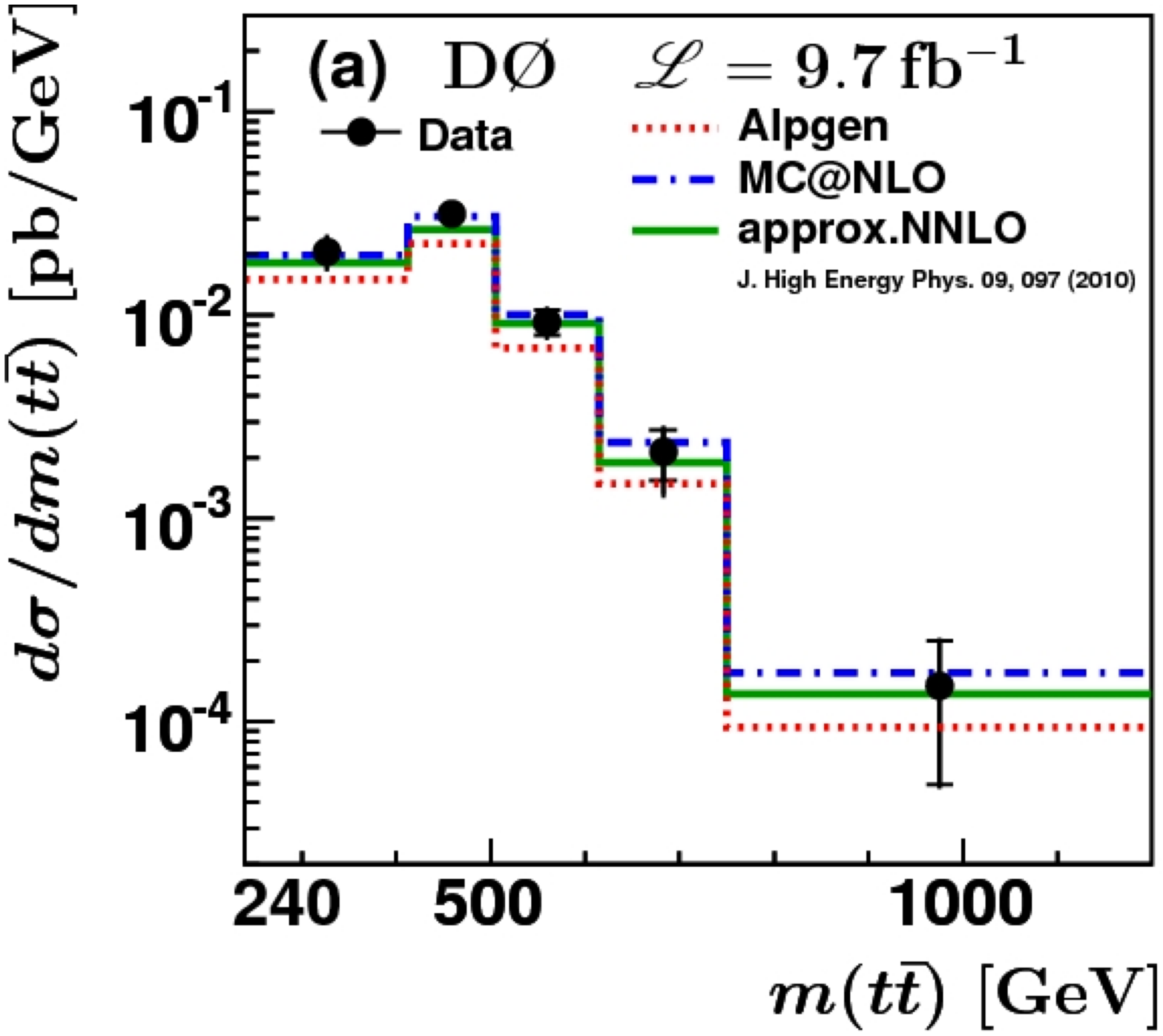}                                                                                                                                                                                                         
\caption{\label{mttbar} Differential $t\bar{t}$ distribution as
  function of $m_{t\bar{t}}$~\protect\cite{diffxsec}.}                                                                                                                                                                          
\end{minipage}\hspace{2pc}%
\begin{minipage}{18pc}                                                                                                                                                                                                                         
\includegraphics[width=20pc]{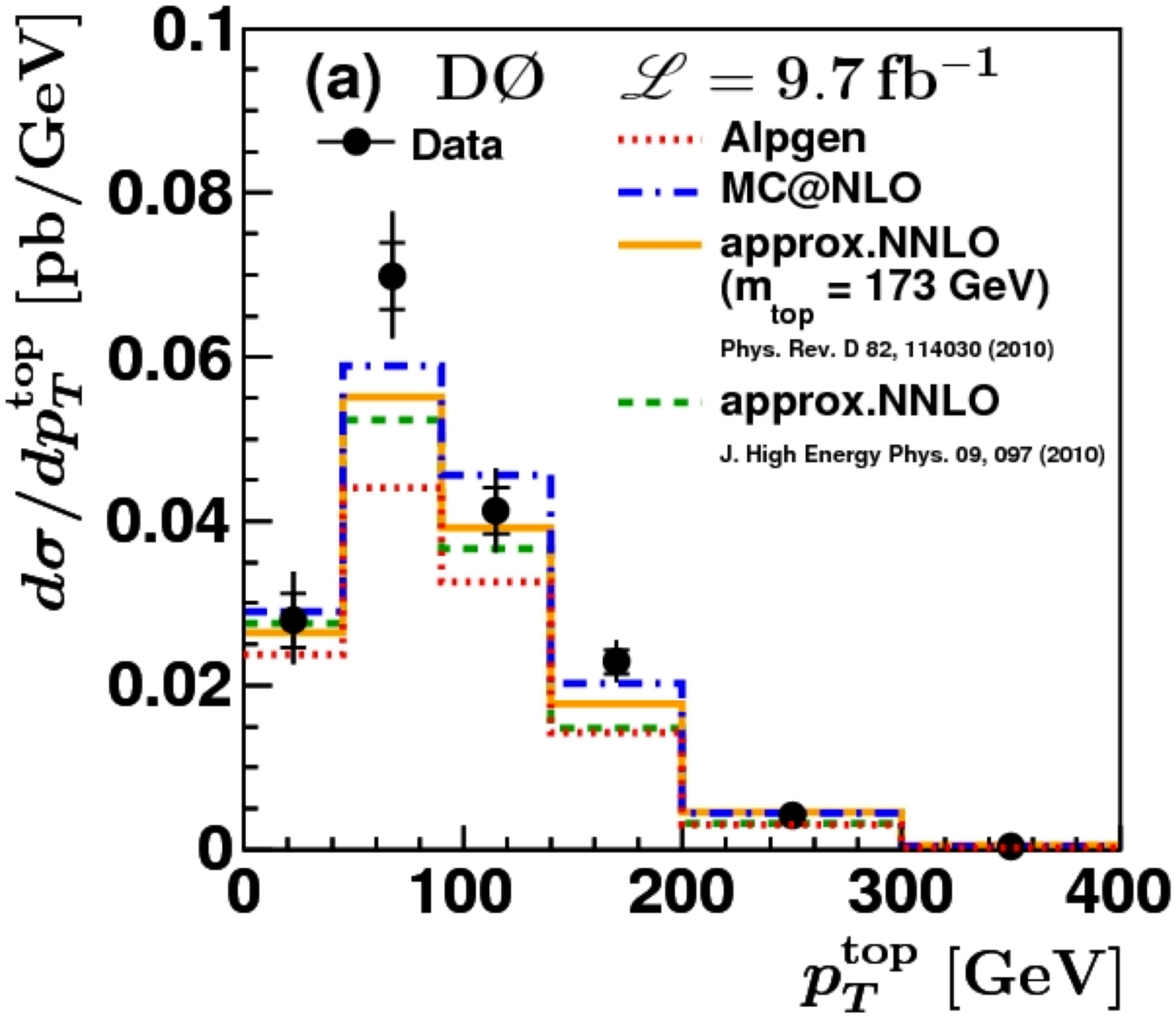}                                                                                                                                                                                                         
\caption{\label{pttop} Differential $t\bar{t}$ distribution as
  function of $p_T^{top}$~\protect\cite{diffxsec}.}                                                                                                                                                                         
\end{minipage}      
\end{figure}

Using the full Run~II data sample of 9.7~fb$^{-1}$, the D0 collaboration recently performed a new measurement of the
$t\bar{t}$ cross section, inclusively as well as differentially as  function of the invariant $t\bar{t}$ mass,
$m_{t\bar{t}}$, the rapidity of the top, $|y^{top}|$, and
 the transverse momentum of the top, $p_T^{top}$~\cite{diffxsec}. For
 the measurement, semileptonic events are analysed. Requiring at least
 four jets in the event, the inclusive cross section yields  $\sigma_{t\bar{t}}=
  8.0 \pm 0.7 {\rm (stat)} \pm 0.6 {\rm (syst)} \pm 0.5 {\rm
    (lumi)}$~pb, in good agreement with the SM prediction. For the
  differential measurement,  the
  $t\bar{t}$ event reconstruction is performed using a constrained
  kinematic fitter, and the distributions are corrected for detector
  and acceptance effects via regularized unfolding. The final
  distributions are defined for parton-level top quarks including
  off-shell effects. The unfolded $t\bar{t}$
  distributions as function of $m_{t\bar{t}}$ and $|p_T^{top}|$ are
  shown in Figure~\ref{mttbar} and Figure~\ref{pttop},
  respectively. The unfolded distributions are compared to approximate
  next-to-next-to-leading order (NNLO) calculations and different
  Monte Carlo generator
  predictions. For all measured distributions, a good agreement between data and the NNLO calculations and
  generator predictions can be seen.

At next-to-leading order (NLO) QCD, interference between different
$q\bar{q}$ diagrams produces a $t\bar{t}$ asymmetry, causing the top
quark to go more likely into the direction of the incoming
quark. Various asymmetries have been studied at the Tevatron, in
particular the
forward-backward asymmetry
$A^{t\bar{t}}_{FB}= \frac {N(\Delta y >0) - N(\Delta y <0)} {N(\Delta
  y >0) + N(\Delta y <0)}$, where $N$ is the number of events with
the difference in top and antitop rapidity $\Delta y$ smaller or
larger than zero, and the leptonic asymmetry $A^{l}_{FB}= \frac
{N(q_l y_l >0) - N(q_l y_l <0)} {N(q_l
  y_l >0) + N(q_l y_l <0)}$, where $q_l$ and $y_l$ are the charge and
rapidity of the lepton from the $W$~boson decay, respectively.

Both CDF and D0 measured the leptonic and the forward-backward
asymmetry in the lepton+jets and dilepton final states. In both cases
the results unfolded to production level are given. Combining
dileptonic and semileptonic events, D0 measures $A_{FB}^{l}= 4.7 \pm
2.3 {\rm (stat)}\pm 1.5 {\rm
  (syst)}$~\%~\cite{d0dilepasym}~\cite{d0asym}, and CDF $A_{FB}^{l}=
9.0^{+2.8}_{-2.6} {\rm (stat+syst)}$~\%~\cite{cdfasym}~\cite{cdfasym2} for the
leptonic asymmetry. These results are in good agreement  with the SM prediction at NLO QCD, including
electroweak (EW) corrections, of $A_{FB}^{l}= 3.8 \pm
0.2$~\%~\cite{bernreutherasym}. The measurement of $A_{FB}^{l}$ as
function of pseudorapidity is also performed,  showing good agreement
with Monte Carlo predictions.  The measurement of $A_{FB}^{t\bar{t}}$
is  done in semileptonic events in D0, resulting in $A_{FB}^{t\bar{t}} = 10.6 \pm 2.7 {\rm (stat)}
\pm 1.3 {\rm (syst)}$~\%~\cite{d0fbasym} and on dileptonic and
semileptonic events in CDF, which yields $A_{FB}^{t\bar{t}} = 16.0 \pm 4.5 {\rm (stat+syst)}$~\%~\cite{cdffbasym2}~\cite{cdffbasym}. Both results are in good agreement with the NNLO SM prediction of 
 $A_{FB}^{t\bar{t}} = 9.5 \pm
 0.7$~\%~\cite{czakonasym}. Recently, D0 performed a new
 measurement of $A_{FB}^{t\bar{t}}$ in the dileptonic final state,
 using the matrix element technique for the $t\bar{t}$ event
 reconstruction. The measured value is  $A_{FB}^{t\bar{t}}  =18.0 \pm
 6.1{\rm (stat)} \pm 3.2 {\rm (syst)}$\%~\cite{d0asymME}. The forward-backward
 asymmetry has also been measured as function of $m_{t\bar{t}}$ and
   $|\Delta y|$. For D0, the resulting distributions are in good
   agreement with Monte Carlo predictions, as well as the CDF
   measurement. 
 


Even though the measured $t\bar{t}$ asymmetries agree with SM
predictions within errors, the experimental value is slightly higher
than the prediction. While the inclusive value is compatible with SM
predictions, especially the differential asymmetry measurements still
do not show a clear picture: both D0 and in particular CDF measure
dependencies of the asymmetry on variables, as for example the
invariant $t\bar{t}$ mass, that has a somewhat higher slope than the
prediction. 
If a non-SM contribution would enhance the
measured  $t\bar{t}$ asymmetry, an enhancement should also be seen for
the  $b\bar{b}$ asymmetry in most models. Both CDF and D0 performed
measurements of the $b\bar{b}$ asymmetry. The CDF collaboration
performed two measurements: an analysis of the $b\bar{b}$ asymmetry at
low invariant $b\bar{b}$ mass, using a soft muon tagging technique to
identify $b$-jets and determine their charge ~\cite{cdfbbasym1}; and
an analysis of the $b\bar{b}$ asymmetry for high   invariant
$b\bar{b}$ mass, where the $b$-jets are identified using lifetime
taggers and the jet charge is determined via a jet charge
algorithm~\cite{cdfbbasym2}. The D0 collaboration considered the asymmetry in
the production of $B^{\pm}$ mesons, using  $B^{\pm} \rightarrow J/\psi
K^{\pm}$ decays~\cite{d0bbasym}. All measurements are consistent with
zero, not indicating any non-SM contributions. 

An important property of the top quark is its mass. The mass is a free
parameter in the SM. The top quark mass, together with the mass of the
$W$ boson, provide a constraint on the Higgs boson mass, and therefore
provide a self-consistency check of the SM. Various methods have been
employed in order to precisely measure the top quark mass, ranging
from template methods to matrix element and ideogram methods. The
matrix element method uses the full event kinematics, therefore being
the most precise method to determine the top mass.  In this method, a
probability is calculated for each event, integrating over leading order matrix elements, folded with parton
distribution functions and transfer functions.
Recently, the D0 collaboration performed a measurement of the top
quark mass in semileptonic events on the full Run~II data sample of
9.7~fb$^{-1}$. In this analysis, the speed of the integration of the
matrix element method has been improved compared to earlier
measurements, allowing the integration of larger MC samples. These
optimizations allowed a more precise and refined study of systematic
uncertainties. To reduce the uncertainty from jet energy
scale, jets from the hadronically decaying $W$ boson are used as an in-situ
constraint. The top quark mass has been measured to be $m_t=174.98 \pm 0.58
{\rm (stat+JES)} \pm 0.49 {\rm (syst)}$~GeV~\cite{d0mtop}~\cite{d0mtop2}, where JES
is the jet energy scale.
Furthermore, the D0 collaboration has released a new measurement of
the top quark mass in dileptonic events, where a neutrino weighting
technique has been applied in order to reconstruct the $t\bar{t}$
event. The JES uncertainty has been reduced by employing the
calibration of JES on hadronically decaying $W$ bosons in semileptonic
$t\bar{t}$ events. The measure top quark mass is $m_t=173.3 \pm 1.4
{\rm (stat)} \pm 0.5 {\rm (JES)} \pm 0.7 {\rm
  (syst)}$~GeV~\cite{d0mtopdilep}. 
In July 2014, a Tevatron top quark mass combination, including the new
D0 top quark mass analysis in semileptonic decays, has been performed,
yielding $m_t=174.34 \pm 0.37
{\rm (stat)} \pm 0.52 {\rm (syst)}$~GeV~\cite{tevmtop}.

\section{Summary}
Even though the LHC is a top quark factory, providing huge amounts of
top quark events, the analysis of Tevatron data is still valuable,
providing complementary information about the
heaviest known elementary particle. Many new
measurements of top quark production and properties have been released
recently by the CDF and  D0 collaborations. Several of these are Tevatron
legacies, as for example the most precise single measurement of the
top quark mass, the final word from the Tevatron  on the forward-backward
asymmetry, and the measurement of a variety of differential $t\bar{t}$
distributions. 

\section*{Acknowledgments}
I thank my collaborators from CDF and  D0
 for their help in preparing the presentation and this
article. I also thank the staffs at Fermilab and
collaborating institutions, and acknowledge the support from the
Helmholtz association.


\section*{References}

\begin{thebibliography}{99}
\bibitem{cdftopdiscovery} F.~Abe {\it et al.}  [CDF Collaboration],
  {\it Phys.\ Rev.\ Lett.\ }  {\bf 74}, 2626 (1995).
\bibitem{d0topdiscovery}  S.~Abachi {\it et al.}  [D0 Collaboration],
  {\it Phys.\ Rev.\ Lett.\ }  {\bf 74}, 2632 (1995).
\bibitem{cdfsingletop}  T. Aaltonen {\it et al.} [CDF  Collaboration],
   {\it Phys. Rev. Lett.} {\bf 103}, 092002 (2009).
\bibitem{d0singletop}  V. M. Abazov {\it et al.} [D0  Collaboration],
  {\it Phys. Rev. Lett. } {\bf 103}, 092001 (2009).
\bibitem{d0tchannel} V. M. Abazov et al. [D0  Collaboration],   {\it
    Phys. Lett. B } {\bf 705}, 313 (2011).
\bibitem{tevschannel} T. Aaltonen {\it et al.} [CDF and D0  Collaborations],   {\it Phys. Rev. Lett.} {\bf 112} , 231803 (2014). 
\bibitem{tevcombinew} T. Aaltonen {\it et al.} [CDF and D0
  Collaborations],    	arXiv:1503.05027, submitted to {\it PRL}.
\bibitem{diffxsec} V. M. Abazov {\it et al.} [D0  Collaboration],
  {\it Phys. Rev. D} {\bf  90}, 092006 (2014).
\bibitem{d0dilepasym} V. M. Abazov et al. [D0  Collaboration],   {\it Phys. Rev. D} {\bf 88}, 112002 (2013).
\bibitem{d0asym} V. M. Abazov {\it et al.} [D0  Collaboration],   {\it Phys. Rev. D} {\bf 90}, 072001 (2014).
\bibitem{cdfasym}  T. Aaltonen {\it et al.} [CDF Collaboration],  {\it Phys. Rev. D} {\bf 88}, 072003 (2013).
\bibitem{cdfasym2} T. Aaltonen {\it et al.} [CDF Collaboration],  {\it
    Phys. Rev. Lett.} {\bf 113}, 042001 (2014).
\bibitem{czakonasym} M. Czakon, P. Fiedler, and A. Mitov, arXiv:1411.3007 (2014).
\bibitem{bernreutherasym}  W.~Bernreuther and Z.~-G.~Si,  {\it Phys.\ Rev.\ D} {\bf 86}, 034026 (2012).
\bibitem{d0fbasym}  V. M. Abazov {\it et al.} [D0  Collaboration],  arXiv:1405.0421 [hep-ex] (submitted to  {\it PRD}). 
\bibitem{cdffbasym2} CDF  Collaboration, CONF-note 11161 (2015).
\bibitem{cdffbasym} T. Aaltonen  {\it et al.} [CDF  Collaboration],
  {\it Phys. Rev. D} {\bf 87}, 092002 (2013).
\bibitem{d0asymME} D0 Collaboration, D0 note 6445-CONF (2014).
\bibitem{cdfbbasym1}  CDF  Collaboration, CONF-note 11156 (2015).
\bibitem{cdfbbasym2}  T. Aaltonen {\it et al.} [CDF Collaboration],
  arXiv:1504.06888 (2015). 
\bibitem{d0bbasym} V. M. Abazov {\it et al.} [D0  Collaboration],
  {\it Phys. Rev. Lett.} {\bf 114}, 051803 (2015).
\bibitem{d0mtop} V. M. Abazov et al. [D0  Collaboration],  {\it
    Phys. Rev. Lett.} {\bf 113}, 032002 (2014).
\bibitem{d0mtop2} V. M. Abazov et al. [D0  Collaboration],  {\it
    Phys. Rev. D} {\bf  91}, 112003 (2015).
\bibitem{d0mtopdilep} D0 Collaboration, D0 note 6463-CONF (2015). 
\bibitem{tevmtop} T. Aaltonen {\it et al.} [CDF and D0  Collaborations],  arXiv:1407.2682 [hep-ex].
\end{thebibliography}
\end{document}